\documentclass[10pt, conference, compsocconf]{IEEEtran}

\usepackage{amsmath}
\usepackage{algorithm}
\usepackage{algorithmic}
\usepackage{graphicx}

\begin{document}

\title{Auto-Tuning Dedispersion for Many-Core Accelerators}

\author{
\IEEEauthorblockN{Alessio Sclocco, Henri E. Bal}
\IEEEauthorblockA{Faculty of Sciences\\
Vrije Universiteit Amsterdam\\
Amsterdam, the Netherlands\\
a.sclocco@vu.nl, h.e.bal@vu.nl}

\and

\IEEEauthorblockN{Jason Hessels, Joeri van Leeuwen}
\IEEEauthorblockA{ASTRON\\
Netherlands Institute for Radio Astronomy\\
Dwingeloo, the Netherlands\\
hessels@astron.nl, leeuwen@astron.nl}

\and

\IEEEauthorblockN{Rob V. van Nieuwpoort}
\IEEEauthorblockA{NLeSC\\
Netherlands eScience Center\\
Amsterdam, the Netherlands\\
r.vannieuwpoort@esciencecenter.nl}
}

\maketitle

\begin{abstract}

Dedispersion is a basic algorithm to reconstruct impulsive astrophysical signals.
It is used in high sampling-rate radio astronomy to counteract temporal smearing by intervening interstellar medium.
% Such smearing is straightforwardly removed if the distance traveled by the signal from the star to the Earth is known.
% In the searches for new celestial objects that radio telescopes often undertake this distance is, however, by definition unknown.
To counteract this smearing, the received signal train must be dedispersed for thousands of trial distances, after which the transformed signals are further analyzed.
This process is expensive on both computing and data handling.
This challenge is exacerbated in future, and even some current, radio telescopes which routinely produce hundreds of such data streams in parallel.
There, the compute requirements for dedispersion are high (petascale), while the data intensity is extreme.
Yet, the dedispersion algorithm remains a basic component of every radio telescope, and a fundamental step in searching the sky for radio pulsars and other transient astrophysical objects.

In this paper, we study the parallelization of the dedispersion algorithm on many-core accelerators, including GPUs from AMD and NVIDIA, and the Intel Xeon Phi.
An important contribution is the computational analysis of the algorithm, from which we conclude that dedispersion is inherently memory-bound in any realistic scenario, in contrast to earlier reports.
We also provide empirical proof that, even in unrealistic scenarios, hardware limitations keep the arithmetic intensity low, thus limiting performance.
We exploit auto-tuning to adapt the algorithm, not only to different accelerators, but also to different observations, and even telescopes.
% Most importantly, we demonstrate that there is no single optimal setting for a platform: our auto-tuning finds that the optimal settings differ for different observation modes on the same architecture.
Our experiments show how the algorithm is tuned automatically for different scenarios and how it exploits and highlights the underlying specificities of the hardware: in some observations, the tuner automatically optimizes device occupancy, while in others it optimizes memory bandwidth.
We quantitatively analyze the problem space, and by comparing the results of optimal auto-tuned versions against the best performing fixed codes, we show the impact that auto-tuning has on performance, and conclude that it is statistically relevant.
% Furthermore, dedispersion represents a class of applications that, due to their high data intensity, are not often implemented on accelerators, pushing the limit of many-core architectures.

\end{abstract}


\section{Introduction} % (fold)
\label{sec:introduction}

Some astronomical sources, such as pulsars, emit millisecond duration, impulsive signals over a wide range of radio frequencies.
As this electromagnetic wave propagates through the ionized material between us and the source, it is dispersed.
This causes lower radio frequencies to arrive progressively later and without correction this results in a loss of signal-to-noise that often makes the source undetectable when integrating over a wide observing bandwidth.
For a fixed interval of frequencies, this dispersion is a non-linear function of the distance between the emitting source and the receiver, that can be reversed by simply shifting in time the signal's lower frequencies.
This process is called \textit{dedispersion}.
% Dedispersion is a basic  algorithm in high-time-resolution radio astronomy, and one of the building blocks of modern radio telescopes like the Low Frequency Array (LOFAR)~\cite{vos2009} and the Square Kilometer Array (SKA)~\cite{carilli2004}.
Dedispersion is a basic  algorithm in high-time-resolution radio astronomy, and one of the building blocks of modern radio telescopes like the Low Frequency Array (LOFAR) and the Square Kilometer Array (SKA).
The amount of processing needed for dedispersion varies per instrument, but can be in the petaflop range.
Hence, it is important to have a high performance, adaptable and portable dedispersion algorithm.

Due to the nature of the problem, however, designing a high performance dedispersion algorithm is far from trivial.
In fact, dispersion can be easily reversed if the distance of the source from the receiver is known in advance, but this is not true when searching for unknown objects in surveys for pulsars or fast-transient sources.
When searching for these celestial objects, the distance is one of the unknowns, and the received signal must be dedispersed for thousands of possible trial distances.
This results in a brute-force search that produces many dedispersed signals, one for each trial distance.
% The data rates, which can already be in the Tb/s range, are thus blown up by an order of magnitude.
Clearly, this search is both computationally and data intensive, and, due to the extremely high data-rate of modern radio telescopes, it must be performed in real-time, since the data streams are too large to store in memory or on disk.
Luckily, all these different searches are independent from each other, and can be performed in parallel.

We aim to achieve high performance by parallelizing this algorithm for many-core accelerators.
Recently, similar attempts have been made by Barsdell et al.~\cite{barsdell2012b} and Armour et al.~\cite{armour2012}.
However, we believe that the performance analysis presented there is not complete.
Moreover, the focus in~\cite{barsdell2012b} and \cite{armour2012} is on specific platforms and observational setups, while in this paper we focus on designing a portable many-core algorithm that can be tuned for different platforms and, more importantly, different radio telescopes and observational setups.
To our knowledge, this is the first attempt at designing a dedispersion algorithm that is not fine tuned for a specific platform or telescope.
Furthermore, even if dedispersion is an inherently parallel algorithm, it is still interesting as it represents a class of applications that, due to their low arithmetic and high data intensity, is not often implemented on accelerators.
We believe that these applications do not only push the limit of many-core architectures, but can also benefit from the higher memory bandwidth that most many-cores provide, compared with traditional CPUs.

We designed and developed a many-core dedispersion algorithm, and implemented it using the Open Computing Language (OpenCL).
Because of its low arithmetic intensity, we designed the algorithm in a way that exposes the parameters controlling the amount of parallelism and possible data-reuse.
In this paper we show how, by auto-tuning these user-controlled parameters, it is possible to achieve high performance on different many-core accelerators, including one AMD GPU (HD7970), three NVIDIA GPUs (GTX 680, K20 and GTX Titan) and the Intel Xeon Phi.
We not only auto-tune the algorithm for different accelerators, but also use auto-tuning to adapt the algorithm to different observational configurations.
In radio astronomy, observational parameters are more variable than the platforms used to run the software, so being able to adapt the algorithm to different observational setups is of major importance.
Furthermore, in this work we measure how much faster a tuned algorithm is compared to every other possible configuration of the parameters, and quantify the statistical difference between optimum and average performance.
Finally, with this work we are able to provide a comparison of modern accelerators based on a real scientific application instead of synthetic benchmarks.

To summarize our contributions, in this paper we: (1) provide an in-depth analysis of the arithmetic intensity of dedispersion, providing analytical evidence and empirical proofs of it being memory-bound, in contrast to earlier claims in the literature; (2) show that by using auto-tuning it is possible to adapt the algorithm to different platforms, telescopes, and observational setups; (3) demonstrate that it is possible to achieve real-time performance using many-core accelerators; (4) quantify the impact that auto-tuning has on performance; and (5) compare different platforms using a real-world scientific application.

% The subsequent parts of this paper are organized as follows.
% Section~\ref{sec:background} provides a brief introduction to the theory of dedispersion and the challenges associated with it.
% Section~\ref{sec:dedispersion_kernel} presents the dedispersion algorithm, introducing our parallel implementation and its optimizations; this section includes the theoretical analysis of dedispersion's arithmetic intensity.
% The experiments used to validate our work are described in Section~\ref{sec:experimental_setup}, while their results are presented in Section~\ref{sec:results_and_discussion}.
% Finally, relevant literature is discussed in Section~\ref{sec:related_works}, and Section~\ref{sec:conclusions} summarizes our conclusions.

% section introduction (end)

\section{Background} % (fold)
\label{sec:background}

Waves traveling through a medium may interact with it in different ways; the result of one of these interactions is called \textit{dispersion}.
The most common example of dispersion comes from the realm of visible light: rainbows.
In the case of a rainbow, the original signal is dispersed when passing through raindrops and its composing frequencies are reflected at different angles.
As a result, we see the components of a whole signal as if they were different ones.

\begin{figure}
\centering
\includegraphics[width=0.60\columnwidth]{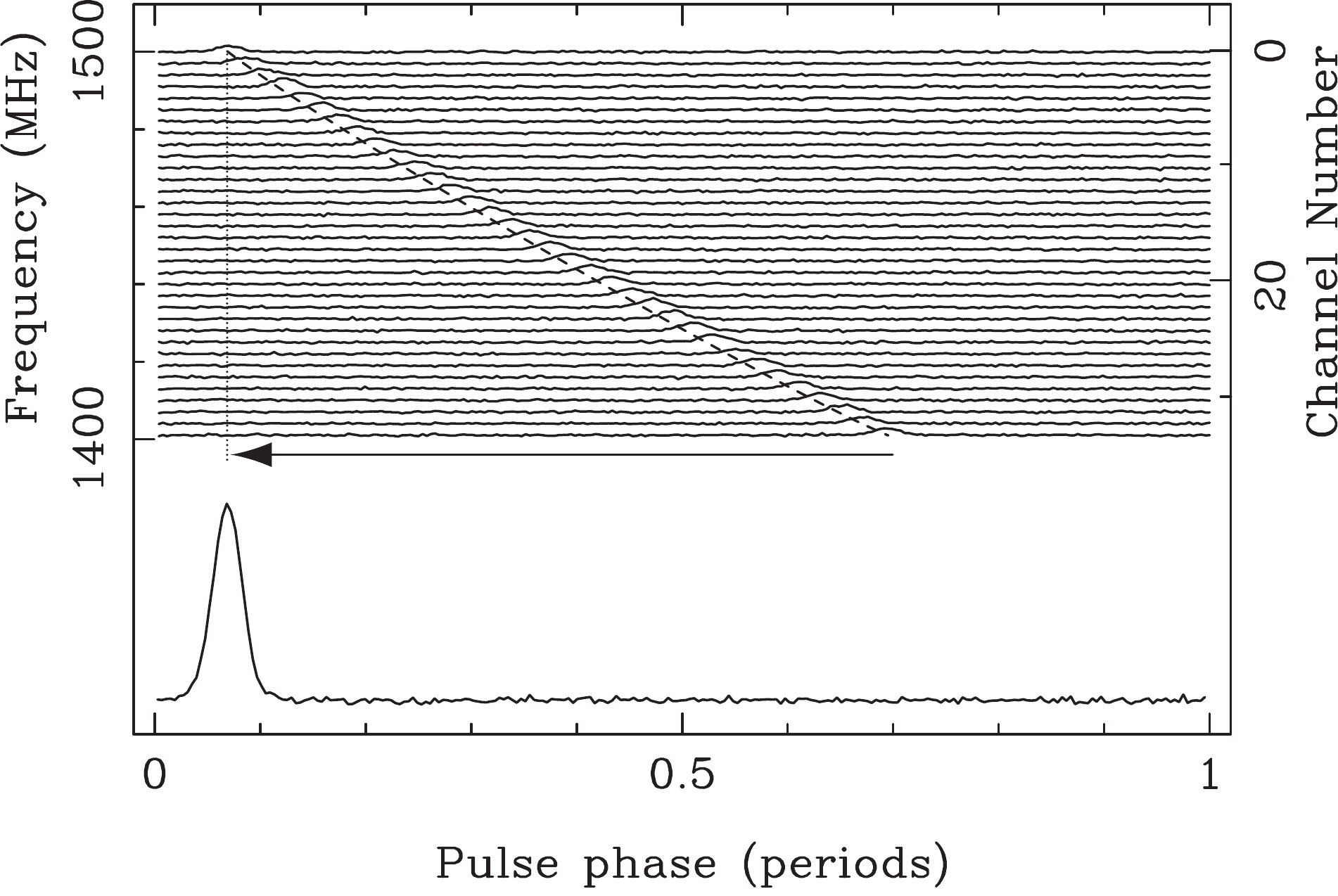}
\caption{The effect of dispersion on a pulsar signal, courtesy of Lorimer and Kramer~\cite{lorimer2005}.}
\label{fig:Dispersion}
\end{figure}

The effect experienced in radio astronomy is similar.
When a celestial source emits an electromagnetic wave, all the frequencies that are part of this emission start their journey together.
However, because of the interaction between the wave itself and the free electrons in the interstellar medium, each of these frequencies is slowed down at a different non-linear rate: lower frequencies experience higher delays.
As a result, when we receive this signal on Earth, we receive the different components at different times, even if they were emitted simultaneously by the source.
Figure~\ref{fig:Dispersion} shows the effect of dispersion on an impulsive radio signal.
The top panel shows the arrival time versus the observing frequency, while the bottom panel shows the dedispersed pulse, which closely approximates the originally emitted signal.

More formally, the highest frequency (i.e. $f_{h}$) of a signal emitted by a specific source at time $t_{0}$ is received on Earth at time $t_{x}$, while all the other frequency components (i.e. $f_{i}$) are received at time $t_{x} + k$.
This delay $k$, measured in seconds, is described by Equation~\ref{eq:Dispersion}; the frequency components of the equation are measured in MHz.

\begin{equation}
\label{eq:Dispersion}
\footnotesize{
	k \approx 4,150 \times DM \times (\frac{1}{f_{i}^{2}} - \frac{1}{f_{h}^{2}})
}
\end{equation}

In this equation, the \textit{Dispersion Measure} (DM) represents the number of free electrons between the source and the receiver.
Assuming a model for the distribution of these electrons along the line-of-sight, this quantity can be interpreted as a measure of the distance between the emitting object and the receiver.
When observing a known object, all the quantities of Equation~\ref{eq:Dispersion} are known, thus the effect of dispersion can be easily reversed.
This process is called \textit{dedispersion} and consists of shifting in time the lower frequencies in order to realign them with the corresponding higher ones, thus reconstructing the original signal.

However, applying this technique in the search for unknown astronomical objects is more difficult because DM is not known a priori.
Therefore, the received signal must be dedispersed for thousands of possible DM values in a brute-force search that produces a new dedispersed signal for every trial DM.
So far, no better approach is known and there are no heuristics available to prune the DM search space. The reason for this is that when the DM is only slightly off, the source signal will be smeared, and the signal strength will drop below the noise floor, becoming undetectable.
It is clear from this that the process of dedispersion is computationally expensive, because every sample in the input signal must be processed for thousands of different trial DMs.
In addition, modern radio telescopes can point simultaneously in different directions by forming different beams.
These beams can be used to survey a bigger part of the sky and discover astronomical sources at a faster rate.
This results in even more input signals to process and greatly increases performance requirements.
However, all trial DMs and beams can be processed independently, thus making it possible to improve the performance of the dedispersion algorithm by means of large-scale parallelization.

% section background (end)

\section{Algorithm and Implementation} % (fold)
\label{sec:dedispersion_kernel}

Dedispersion is the process of reversing the effects of dispersion, as described in Section~\ref{sec:background}.
We first describe the sequential dedispersion algorithm in more detail, and analyze its complexity in Section~\ref{sub:sequential_algorithm}. We then present our parallel implementation and its optimizations in Section~\ref{sub:parallelization}.

\subsection{Sequential Algorithm} % (fold)
\label{sub:sequential_algorithm}

The input of this algorithm is a channelized time-series, i.e. a time-series with each frequency channel represented as a separate component.
The time-series is represented as a $c \times t$ matrix, where $c$ is the number of frequency channels and $t$ is the number of time samples necessary to dedisperse one second of data at the highest trial DM.
This number, $t$, is always a multiple of the number of samples per second.
The output is a set of dedispersed time-series, one for each trial DM, and it is represented as a $d \times s$ matrix, where $d$ is the number of trial DMs and $s$ is the number of samples per second. During dedispersion, the frequency channels are typically integrated to reduce the data rate. 
Every data element in these matrices is represented using a single precision floating point number.
% To provide a quantitative example, the Apertif on Westerbork telescope~\cite{verheijen2008} will receive 36~GB/s in input, and produce 72~GB/s of dedispersed data.
To provide a quantitative example, the Apertif system on the Westerbork telescope will receive 36~GB/s in input, and produce 72~GB/s of dedispersed data.

The sequential pseudocode for the dedispersion algorithm is shown in Algorithm~\ref{alg:Dedispersion}.
Even if the algorithm looks trivially parallel, it is very data-intensive, so achieving high performance is not trivial.
To simplify the discussion, and without losing generality, in this paper we describe the case in which there is a single input beam, but all results can be applied to the case of multiple beams.
The algorithm consists of three nested loops, and every output element is the sum of $c$ samples: one for each frequency channel.
Which samples are part of each sum depends on the applied delay (i.e. $\Delta$) that, as we know from Equation~\ref{eq:Dispersion}, is a non-linear function of frequency and DM.
% Figure~\ref{fig:DelayApertif} provides an example of the shape of this delay function for different DM values.
These delays can be computed in advance, so they do not contribute to the algorithm's complexity.
Therefore, the complexity of this algorithm is $O(d \times s \times c)$.

\begin{algorithm}
\caption{Pseudocode of the dedispersion algorithm.}
\label{alg:Dedispersion}
\footnotesize{
\begin{algorithmic}
\FOR{dm = 0 $\to$ d}
  \FOR{sample = 0 $\to$ s}
    \STATE{dSample = 0}
    \FOR{channel = 0 $\to$ c}
      \STATE{dSample += input[channel][sample + $\Delta$(channel, dm)]}
    \ENDFOR
    \STATE{output[dm][sample] = dSample}
  \ENDFOR
\ENDFOR
\end{algorithmic}
}
\end{algorithm}

% \begin{figure}
% \centering
% \includegraphics[width=0.65\columnwidth]{delayApertif}
% \caption{An example of delays, in samples, for different DMs.}
% \label{fig:DelayApertif}
% \end{figure}

In the context of many-core accelerators, there is another algorithmic characteristic that is of great importance: \textit{Arithmetic Intensity} (AI), i.e. the ratio between the performed floating point operations and the number of bytes accessed in memory.
The AI is extremely important, because in many-core architectures the gap between computational capabilities and memory bandwidth is wide, thus a high AI is a prerequisite for high performance \cite{williams2009}.
Unfortunately, Algorithm~\ref{alg:Dedispersion} shows that dedispersion's AI is inherently low, as there is only one floating point operation for every input element loaded from global memory.
A bound for the AI of dedispersion is presented in Equation~\ref{eq:DedispersionAI}, where $\epsilon$ represents the effect of accessing the delay table and writing the output.

\begin{equation}
\label{eq:DedispersionAI}
\footnotesize{
  AI = \frac{1}{4 + \epsilon} < \frac{1}{4}
}
\end{equation}

The low AI of Equation~\ref{eq:DedispersionAI} identifies dedispersion as a memory-bound algorithm on most architectures, thus the performance of this algorithm is limited not by the computational capabilities of the architecture used to execute it, but by its memory bandwidth.
A way to increase dedispersion's AI, thus improving performance, is to reduce the number of reads from global memory by implementing some form of data-reuse.
Analysis of Algorithm~\ref{alg:Dedispersion} suggests that some data-reuse may indeed be possible.
Given that the time dimension is represented with discrete samples, it may happen that, for some frequencies, the delay is the same for two close DMs, $dm_{i}$ and $dm_{j}$, so that $\Delta(c, dm_{i}) = \Delta(c, dm_{j})$.
In this case, the same input element can be used to compute two different sums, thus offering the opportunity for data-reuse and an improved AI.
% Figure~\ref{fig:DelayApertifZoom} shows how the delays at high frequencies may be the same for different DMs.
If this data-reuse is exploited, we can compute a new upper bound for the AI; this new upper bound is presented in Equation~\ref{eq:DedispersionAIp}.

\begin{equation}
\label{eq:DedispersionAIp}
\footnotesize{
  AI < \frac{1}{4 \times (\frac{1}{d} + \frac{1}{s} + \frac{1}{c})} 
}
\end{equation}

The bound from Equation~\ref{eq:DedispersionAIp} goes towards infinity.
It may be tempting to overestimate this theoretical result and believe that, by exploiting data-reuse, dedispersion's AI can be increased enough to make the algorithm itself compute-bound.
However, we found this is not the case in any realistic scenario.
To approximate this upper bound, data-reuse should be possible for every combination of DMs and frequencies, but the delay function is not linear, and delays diverge rapidly at lower frequencies.
This means that, in realistic scenarios, there will never be enough data-reuse to approach the upper bound of Equation~\ref{eq:DedispersionAIp}.
Moreover, using the same delay for every combination of DMs and frequencies would produce the same result for every dedispersed time-series, thus making the computed results useless (i.e. the DM step is too small).
Therefore we conclude that, even if data-reuse is possible, it depends on parameters like DM values and frequencies that cannot be controlled and this makes the upper bound on the algorithm's AI presented in Equation~\ref{eq:DedispersionAIp} not approachable in any realistic scenario.
In this conclusion we differ from previous literature like \cite{barsdell2010} and \cite{barsdell2012b}. We will prove our claim with experimental results in Section~\ref{sub:data_reuse_and_performance_limits}.

% \begin{figure}
% \centering
% \includegraphics[width=0.65\columnwidth]{delayApertifZoom}
% \caption{Possible data reuse for different DMs.}
% \label{fig:DelayApertifZoom}
% \end{figure}

% subsection sequential_algorithm (end)

\subsection{Parallelization} % (fold)
\label{sub:parallelization}

The first step in parallelizing the dedispersion algorithm for many-cores is to determine how to divide the work among different threads and how to organize them; in this work we use OpenCL as our many-core implementation framework, thus we utilize OpenCL terminology when referring to threads and their organization.
From the description of the sequential dedispersion algorithm and its data structures we can identify three main dimensions: DM, time and frequency.
Of these three dimensions, DM and time are the ones that are independent from each other.
Moreover, they also lack internal dependencies, thus every two time samples or DMs can be computed independently of each other.
These properties make the two dimensions ideal candidates for parallelization, avoiding any inter- and intra-thread dependency.
In our implementation, each OpenCL work-item (i.e. thread) is associated with a different (DM, time) pair and it executes the innermost loop of Algorithm~\ref{alg:Dedispersion}.
An OpenCL work-group (i.e. a group of threads) combines work-items that are associated with the same DM, but with different time samples.

This proposed organization has another advantage other than thread independence: it simplifies the memory access pattern making it possible to have coalesced reads and writes.
In many architectures, memory accesses generated by different threads are coalesced if they refer to adjacent memory locations, so that the different small requests are combined together in one bigger operation.
Coalescing memory accesses is a well-known optimization, and it is usually a performance requisite for many-core architectures, especially in case of memory-bound algorithms like dedispersion.
In our implementation, consecutive work-items in the same work-group write their output element to adjacent, and aligned, memory locations, thus accessing memory in a coalesced and cache-friendly way.

Reads from global memory are also coalesced but, due to the shape of the delay function, are not always aligned.
This lack of alignment can be a problem because, on most architectures, memory is transferred in cache-lines, and unaligned accesses can require the transfer of more than a cache-line, thus introducing memory overhead.
In the case of dedispersion, the delay function is part of the algorithm and cannot be modified, but, if a cache-line contains the same number of element as the number of work-items executed in parallel by a Compute Unit (CU), then the memory overhead is at most a factor two.
Luckily, this property holds for most many-cores.
A factor two overhead may have a big impact on performance, but this worst-case scenario applies only if the number of work-items per work-group is the same as the number of work-items executed in parallel by a CU.
In the case of work-groups with more work-items, there is a high chance that the unnecessarily transferred elements will be accessed by other work-items in the near future, thus compensating the introduced overhead with a basic form of prefetching.

We affirmed, in Section~\ref{sec:dedispersion_kernel}, that data-reuse is possible when, for two different DMs, the delay corresponding to a particular frequency channel is the same.
To exploit this data-reuse, and thus increase the algorithm's AI, the parallel algorithm needs to be slightly modified to compute more than one DM per work-group.
So, the final structure of our many-core dedispersion algorithm consists of two-dimensional work-groups.
In this way a work-group is associated with more than one DM, so that its work-items either collaborate to load the necessary elements from global to \textit{local memory}, a fast memory area that is shared between the work-items of a same work-group, or rely on the cache, depending on the architecture.
Therefore, when the same element is needed by more than one work-item, the accesses to memory are reduced. 
There is no penalty introduced with this organization, and everything discussed so far still holds because the one-dimensional configuration is just a special case of the two-dimensional one.
Work-items can also be modified to compute more than one output element, thus increasing the amount of work per work-item to hide memory latency.
Accumulators are kept in registers by the work-items, a further optimization to reduce accesses to global memory.

The general structure of the algorithm can be specifically instantiated by configuring four user-controlled parameters.
Two parameters are used to control the number of work-items per work-group in the time and DM dimensions, regulating the amount of available parallelism.
The other two parameters are used to control the number of elements a single work-item computes, also in the time and DM dimensions, regulating the amount of work per work-item.
The source code implementing a specific instance of the algorithm is generated at run-time, after the configuration of these four parameters.

% subsection parallelization (end)

% section dedispersion_kernel (end)

\section{Experimental Setup} % (fold)
\label{sec:experimental_setup}

In this section we describe how the experiments are carried out and all the necessary information to replicate them.
We start by describing the many-core accelerators we used, the software configuration of our systems and the observational setups.
We then describe the specificities of each of the three experiments that are the focus of this paper.
Table~\ref{tab:Platforms} contains a list of the many-core accelerators we used in this work, and reports some basic details for each of them.
In particular, the table shows each platform's number of \textit{Compute Elements} (CEs), peak performance and peak memory bandwidth.

\begin{table}
\centering
\footnotesize{
\begin{tabular}{ | l | r | r | r | }
\hline
\textbf{Platform} & \textbf{CEs} & \textbf{GFLOP/s} & \textbf{GB/s} \\
\hline
AMD HD7970 & $64 \times 32$ & 3,788 & 264 \\
\hline
Intel Xeon Phi 5110P & $2 \times 60$ & 2,022 & 320 \\
\hline
% NVIDIA GTX 480 & $32 \times 15$ & 1,344 & 177 \\
% \hline
NVIDIA GTX 680 & $192 \times 8$ & 3,090 & 192 \\
\hline
NVIDIA K20 & $192 \times 13$ & 3,519 & 208 \\
\hline
NVIDIA GTX Titan & $192 \times 14$ & 4,500 & 288 \\
\hline
% Intel Xeon E5-2620 & $6 \times 1$ & 192 & 42 \\
% \hline
\end{tabular}
}
\caption{Characteristics of the used many-core accelerators.}
\label{tab:Platforms}
\end{table}

We run the same code on every many-core accelerator; the source code is implemented in C++ and OpenCL.
The OpenCL runtime used for the AMD HD7970 GPU is the AMD APP SDK 2.8, the runtime used for the Intel Xeon Phi is the Intel OpenCL SDK XE 2013 R3 and the runtime used for the NVIDIA GPUs is NVIDIA CUDA 5.0; the C++ compiler  is version 4.4.7 of the GCC.
% The accelerators are installed in different nodes of the Distributed ASCI Supercomputer 4 (DAS-4)~\footnote{http://www.cs.vu.nl/das4/}.
The accelerators are installed in different nodes of the Distributed ASCI Supercomputer 4 (DAS-4).
DAS-4 runs CentOS 6, with version 2.6.32 of the Linux kernel.
In all experiments, the algorithm is executed ten times, and the average of these ten executions is used for the measurements.
Dedispersion is always used as part of a larger pipeline, so we can safely assume that the input is already available in the accelerator memory, and the output is kept on device for further processing.
There is no need in this scenario to measure data-transfers over the PCI-e bus.

% TODO gflops for apertif cluster, lofar is largest radio telescope in the world
The experiments are performed in two different observational setups.
These setups are based on the characteristics of two radio telescopes operated by the \textit{Netherlands Institute for Radio Astronomy} (Astron): LOFAR and the Apertif system on Westerbork.
In our Apertif setup, the time resolution is 20,000 samples per second.
The bandwidth is 300~MHz, divided in 1,024 frequency channels of 0.29~MHz each; the lowest frequency is 1,420~MHz, the highest frequency is 1,720~MHz.
In our LOFAR setup, the time resolution is higher, 200,000 samples per second, but the bandwidth is lower, 6~MHz divided in 32 frequency channels of 0.19~MHz each; the lowest frequency is 138~MHz, the highest frequency is 145~MHz.
In both setups, the first trial DM is 0 and the increment between two successive DMs is 0.25 $pc/cm^{3}$.

The differences between the two setups are important because they highlight different aspects of the algorithm.
Specifically, the Apertif setup is more computationally intensive, as it involves 20~MFLOP per DM, three times more than the LOFAR setup with just 6~MFLOP per DM.
However, the frequencies in Apertif are much higher than in LOFAR, thus the delays are smaller and there is more available data-reuse.
Therefore, we represent two different and complementary scenarios: one that is more computationally intensive, but offers more possible data-reuse, and one that is less computationally intensive, but precludes almost any data-reuse.

\subsection{Auto-Tuning} % (fold)
\label{sub:exp_auto_tuning}

Our first experiment consists of the auto-tuning of the four algorithm parameters described in Section~\ref{sec:dedispersion_kernel}.
The goal of this experiment is to find the optimal configuration of these parameters for the five many-core accelerators, in both observational setups.
Without auto-tuning we have no a priori knowledge that can guide us into selecting optimal configurations, thus the need for this experiment.
Moreover, we are interested in understanding how these configurations differ, and if a generic configuration can be identified.

The algorithm is executed for every meaningful combination of the four parameters, on every accelerator, and for both operational setups.
A configuration is considered meaningful if it fulfills all the constraints posed by a specific platform, setup and input instance.
An input instance is defined by the number of DMs that the algorithm has to dedisperse.
In this experiment we use 12 different input instances, each of them associated with a power of two between 2 and 4,096.
Due to memory constraints, some platforms may not be able to compute results for all the input instances.
The optimal configuration is chosen as the one that produces the highest number of single precision floating point operations per second.
The output of this experiment is a set of tuples representing the optimal configuration of the algorithm's parameters; there is a tuple for every combination of platform, observational setup and input instance.

% subsection exp_auto_tuning (end)

\subsection{Impact of Auto-Tuning on Performance} % (fold)
\label{sub:exp_impact_of_auto_tuning}

Our second experiment measures the performance of the previously tuned dedispersion algorithm.
The goal of this experiment is twofold.
On the one hand, we want to show performance and scalability of the tuned algorithm, and evaluate the suitability of many-core accelerators to perform this task for current and future radio telescopes.
On the other hand, we want to provide a scientific evaluation of the impact that auto-tuning has on the performance of dedispersion.
The byproduct of this experiment is a comparison of different platforms in the context of radio astronomy.

The algorithm is executed on every accelerator, for each combination of observational setup and input instance.
For each of these runs, the optimal configuration is used (which we found in the first experiment).
The metric used to express performance is the number of single precision floating point operations per second.
To quantify the impact of auto-tuning on performance we present the signal-to-noise ratio of the optimums.
% TODO one sentence on why? We aim to show that the effect of autotuning is statisically relevant.

% subsection exp_impact_of_auto_tuning (end)

\subsection{Data-reuse and Performance Limits} % (fold)
\label{sub:exp_data_reuse_and_performance_limits}

Our third experiment consists of measuring the performance of dedispersion in a situation in which optimal data-reuse is made possible by artificially modifying the delays.
The goal of this experiment is to show how much the observational setup affects performance.
Moreover, we want to provide empirical proof that, even with perfect data-reuse, it would not be possible to achieve high AI because of the limitations of real hardware, in contrast to what reported in literature.

This experiment performs the same steps of experiments 1 and 2, with the only difference that all the DM values are the same: the only DM value used is 0, thus there are no shifts.
With no shifts applied, perfect data-reuse becomes theoretically possible as every dedispersed time-series is exactly the same and uses exactly the same input.
We tune the algorithm in this modified setup, and then measure the performance obtained using the optimal configurations.
The obtained results are compared with the ones obtained in experiment 2 to measure the impact that data-reuse has on performance, and understand the limitations that real hardware poses on performance.

% subsection exp_data_reuse_and_performance_limits (end)

% section experimental_setup (end)

\section{Results and Discussion} % (fold)
\label{sec:results_and_discussion}

In this section we present the results of the experiments described in Section~\ref{sec:experimental_setup}.
For every experiment, we first report the measured results, and then discuss them.
When results are different from what one would expect, we provide a further explanation.
After each experiment, we provide a short summary of the main findings.
To conclude, Section~\ref{sub:discussion} contains some additional performance comparisons.

\subsection{Auto-Tuning} % (fold)
\label{sub:auto_tuning}

\begin{figure}
\centering
\includegraphics[width=0.65\columnwidth]{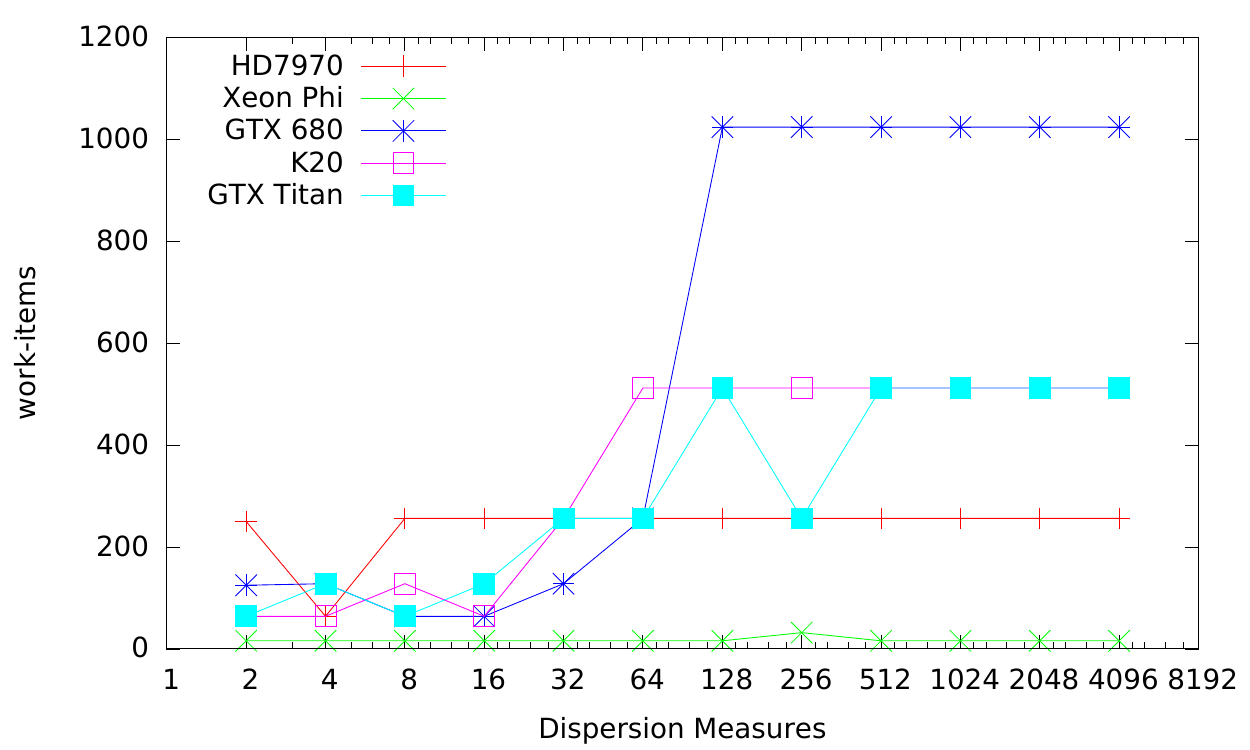}
\caption{Tuning the number of work-items per work-group, Apertif.}
\label{fig:TuningApertifWI}
\end{figure}

In this section we present the results of the auto-tuning experiment described in Section~\ref{sub:exp_auto_tuning}.
We begin by examining the results for the Apertif case.
Figure~\ref{fig:TuningApertifWI} shows the optimal number of work-items per work-group identified by auto-tuning.
The GTX 680 requires the highest number of work-items (1,024), the Xeon Phi requires the lowest (16), and the other three platforms require between 256 and 512 work-items per work-group.
The first noticeable pattern is that the optimal configuration, for each platform, is more variable with smaller input instances and then becomes more stable for bigger instances.
The reason is the amount of work made available by every instance.
In fact, smaller instances expose less parallelism and also have a smaller optimization space associated.
When the number of DMs increases, so does the amount of work and the optimization space; thus, after a certain threshold, there are no more sensible variations in the number of work-items necessary for optimal performance.

\begin{figure}
\centering
\includegraphics[width=0.65\columnwidth]{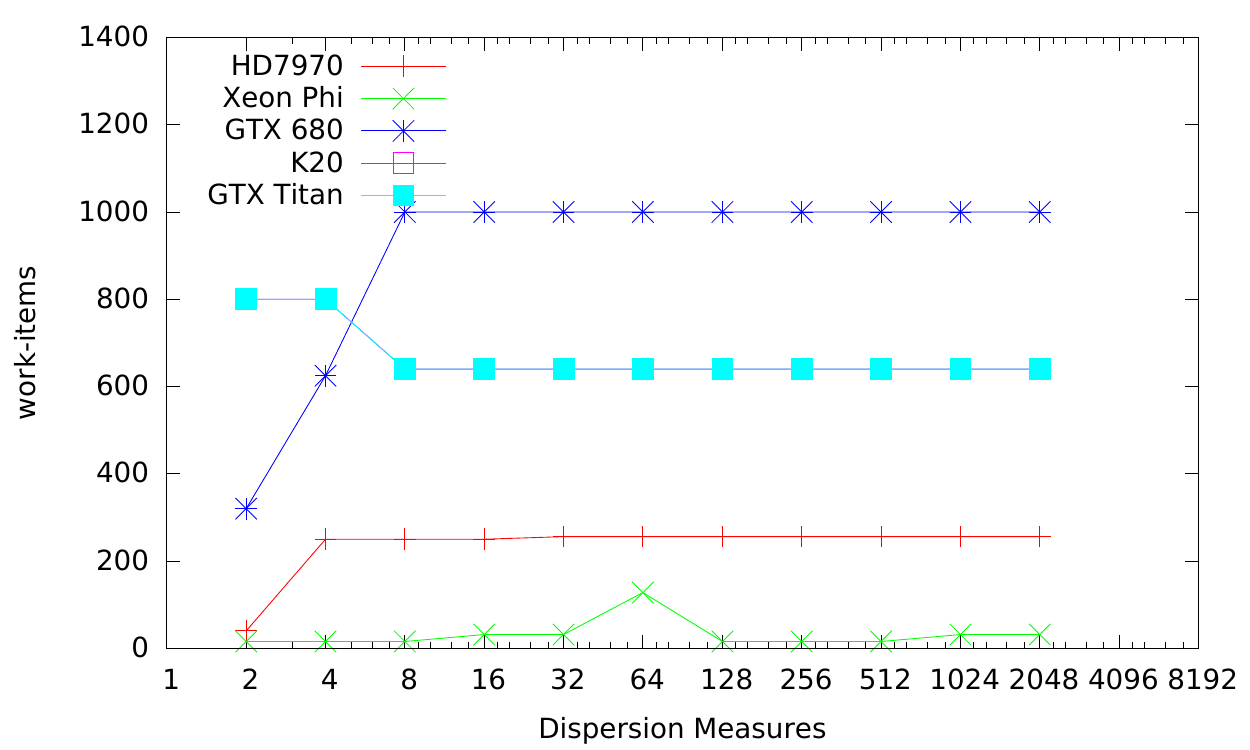}
\caption{Tuning the number of work-items per work-group, LOFAR.}
\label{fig:TuningLOFARWI}
\end{figure}

Figure~\ref{fig:TuningLOFARWI} presents the results of the experiment using the LOFAR observational setup.
These results are similar to the Apertif setup, but not completely.
The first difference that can be identified is that the behavior is more stable: the platforms already reach their optimum with smaller inputs.
The reason is that the LOFAR setup has less available data-reuse, so it is easier to find the optimum because memory-reuse is not increased, even if there are more DMs to compute.
The GTX 680 remains the platform that needs the highest number of work-items (i.e. 1,000) to achieve its top performance, and the Xeon Phi still needs the lowest.
The HD7970 maintains its optimum at 256 work-items per work-group, its hardware limit for the number of work-items per work-group.
Titan and the K20 can be clustered at a mid-range interval.

Overall, the two setups seem similar and the different platforms behave coherently.
However, in Section~\ref{sec:dedispersion_kernel} we pointed out that the total number of work-items per work-group is determined by the interaction of two parameters, with each work-group organized as a two-dimensional structure.
Therefore, what looks like the same result may in fact be obtained in a complete different way.
As an example, the GTX 680 seems to find the same optimum in both setups: 1,024 work-items for Apertif and 1,000 for LOFAR.
However, even if the numbers are similar, they are the results of two different configurations: for Apertif the work-group is organized as a square matrix of $32 \times 32$ work-items, while for LOFAR the matrix is rectangular and composed of $250 \times 4$ work-items.
Taking into account the meaning of these matrices, described in Section~\ref{sub:parallelization}, it is possible to see that in the Apertif setup, where more data-reuse is potentially available, the auto-tuning identifies a configuration that exploits this data-reuse intensively, while in the LOFAR setup, where less reuse is available, the optimal configuration relies less on reuse and more on the device occupancy.
This result is clearly important because it shows not just that the algorithm can be adapted to different observational setups, but also that there is no single optimal configuration.

\begin{figure}
\centering
\includegraphics[width=0.65\columnwidth]{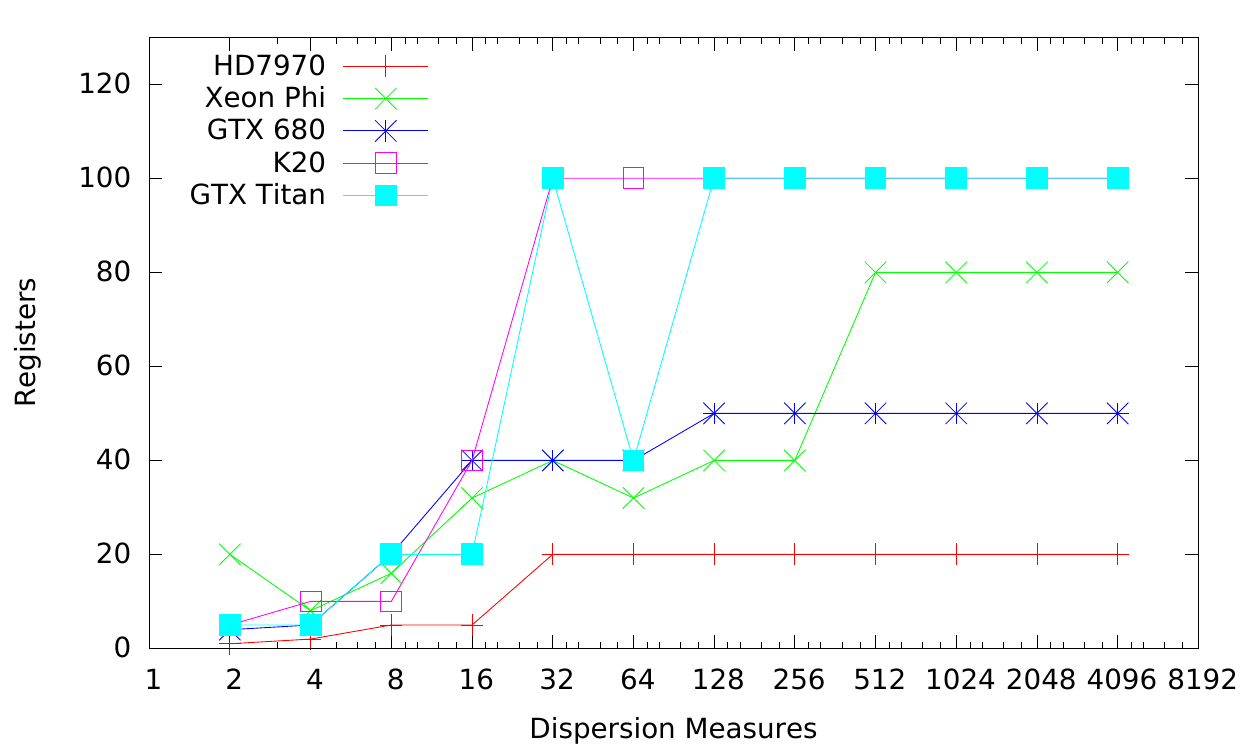}
\caption{Tuning the number of registers per work-item, Apertif.}
\label{fig:TuningApertifReg}
\end{figure}

The combination of the other two parameters of our dedispersion algorithm affects the number of registers that a work-item uses, thus affecting the amount of work that a single work-item is responsible for.
Figure~\ref{fig:TuningApertifReg} illustrates the results for the Apertif setup.
In the figure, the K20 and GTX Titan top the group, followed by Xeon Phi and GTX 680, while the last stand is taken by the HD7970.
In these results we can spot another architectural property: K20 and Titan have the highest number of potentially available registers per work-item among our accelerators, and auto-tuning exploits this property.
In fact, combining the K20 and Titan's results from both Figures~\ref{fig:TuningApertifWI} and \ref{fig:TuningApertifReg}, it is possible to see that the optimal configuration found by auto-tuning for these accelerators is to have fewer work-items than the maximum, but with more work associated.

\begin{figure}
\centering
\includegraphics[width=0.65\columnwidth]{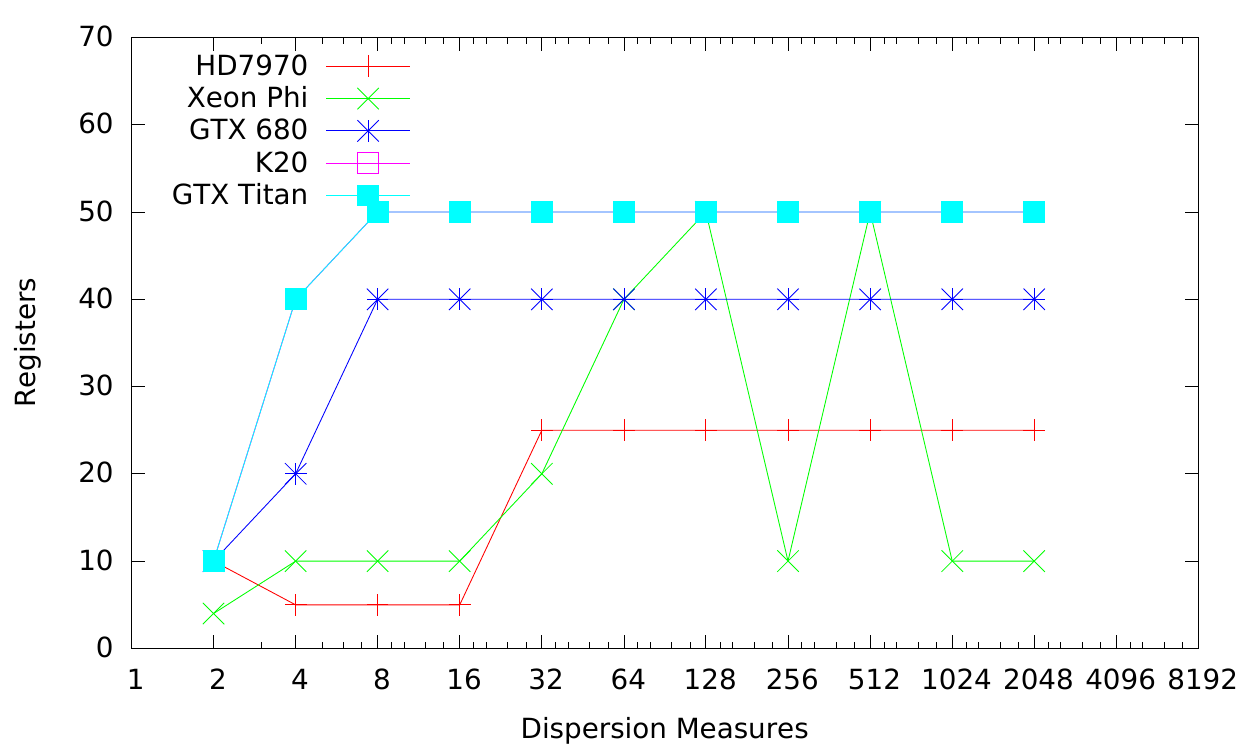}
\caption{Tuning the number of registers per work-item, LOFAR.}
\label{fig:TuningLOFARReg}
\end{figure}

Figure~\ref{fig:TuningLOFARReg} presents the results for the LOFAR setup.
The HD7970 uses the lowest number of registers, keeping its work-items lighter than the other platforms, while the Xeon Phi uses a highly variable number of them.
K20 and GTX Titan use the largest number of registers, but the distance from the GTX 680 is less pronounced than in the previous setup.
In this case, the optimum for these two platforms trades registers for work-items, relying more on parallelism than on heavyweight work-items.
Moreover, by analyzing the composition of these results we can observe once more how auto-tuning enhances the adaptability of dedispersion.
In fact, the optimal register configuration for K20 and Titan is $25 \times 4$ in the Apertif setup, and $25 \times 2 $ in the LOFAR setup: also in this case, the configurations reflect the amount of possible data-reuse and show how auto-tuning makes it possible to adapt the algorithm to different scenarios.

To summarize the results of this experiment, we identified the optimal configurations of the four parameters of our dedispersion algorithm for five different many-core accelerators and two different observational setups.
We identified the interactions between the configurations of these parameters, and noticed how auto-tuning provides the algorithm with the ability to adapt to different scenarios.
We observed that, in general, NVIDIA GPUs require more resources, either work-items per work-group or registers, to achieve their top performance, compared with the AMD and Intel accelerators.
We believe that it would not be possible to identify the optimal configurations a priori and that auto-tuning is the only feasible way to properly configure the dedispersion algorithm, because of its number of parameters, their interaction with each other, and the impact that they have on a fundamental property like the algorithm's AI.
Moreover, optimal configurations are platform and observation specific, and there is no general configuration that can be used in every situation.

% subsection auto_tuning (end)

\subsection{Impact of Auto-Tuning on Performance} % (fold)
\label{sub:impact_of_auto_tuning}

\begin{figure}
\centering
\includegraphics[width=0.65\columnwidth]{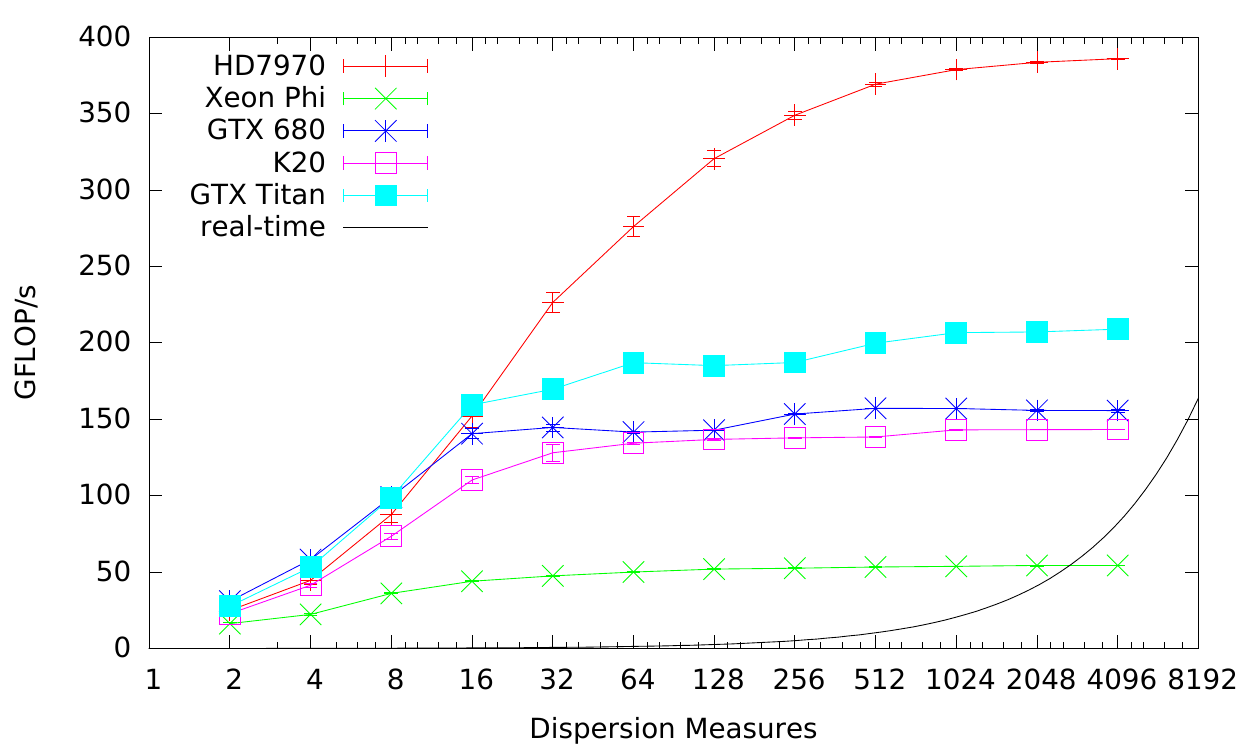}
\caption{Performance of auto-tuned dedispersion, Apertif (higher is better).}
\label{fig:PerfApertif}
\end{figure}

In this section we present and analyze the results of the experiment described in Section~\ref{sub:exp_impact_of_auto_tuning}.
We start by introducing the results of the Apertif setup.
Figure~\ref{fig:PerfApertif} shows the performance achieved by the auto-tuned dedispersion on the various many-core accelerators that we used in this paper.
All the platforms show a similar behavior, with performance increasing with the dimension of the input instance up to a maximum, and plateauing afterwards.
The first noticeable result is that the tuned algorithm scales better than linearly up to this maximum, and then scales linearly.
The accelerators can be clustered in three groups: the HD7970 achieves the highest performance, the Xeon Phi the lowest, and the three NVIDIA GPUs, close to each other in performance, sit in the middle.
On average the HD7970 is 2 times faster than the NVIDIA GPUs, and 7.5 times faster than the Xeon Phi.
%The low Xeon Phi performance is partially due to its immature OpenCL implementation.

% \textbf{The relative order among the three NVIDIA GPUs is unexpected, with the consumer grade GPUs performing better than the K20.
% However, one of the reasons for this experiment is to show the performance achieved by modern accelerators running real-world radio astronomy algorithms, and unexpected results are important to identify the architectures that are best suited for this science.
% To explain the relative order among the NVIDIA GPUs we have to look at the number of CUs and peak memory bandwidth of each GPU; this information is available in Table~\ref{tab:Platforms}.
% The GTX Titan executes one work-group per CU, thus achieving a bandwidth per CU of 11.8~GB/s.
% GTX 680 and K20 execute two work-groups per CU, so their memory bandwidths per CU are 12 and 8~GB/s respectively.
% The memory bandwidth per CU explains the relative order among the NVIDIA GPUs, and even the fact that the consumer cards are closer in performance to each other than to the K20.}

\begin{figure}
\centering
\includegraphics[width=0.65\columnwidth]{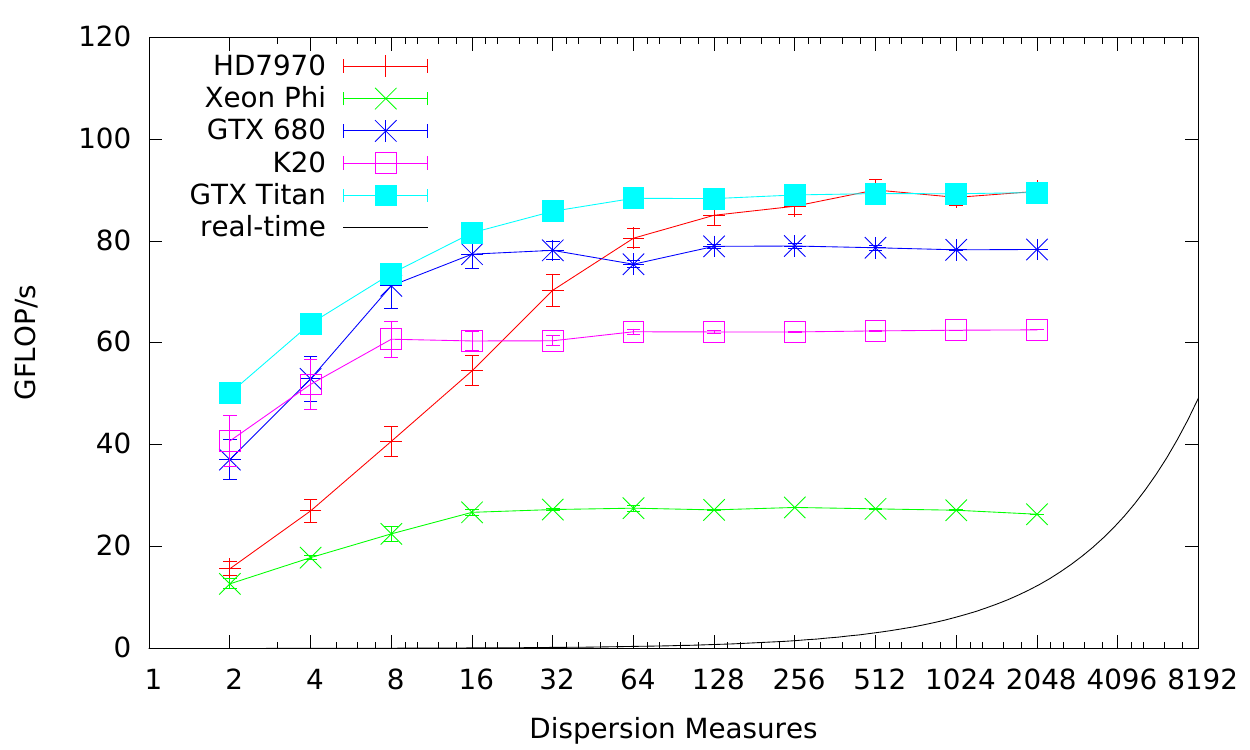}
\caption{Performance of auto-tuned dedispersion, LOFAR (higher is better).}
\label{fig:PerfLOFAR}
\end{figure}

The results for the LOFAR scenario are different, as can be seen from Figure~\ref{fig:PerfLOFAR}.
The first difference is in terms of absolute performance, with performance for LOFAR being lower than in the case of Apertif.
The reason for lower performance can be found in the fact that in this setup there is less available data-reuse, thus the algorithm's AI is lower.
% However, performance requirements of this setup are also lower.
% Therefore, a lower performance in absolute terms does not affect scalability or the satisfaction of the real-time constraint, as can also be seen from the figure.
The other difference is that the GPUs are closer in performance to each other, with the HD7970 and the GTX Titan achieving the higher performance.
In fact, with less data-reuse available, the discriminant for performance is memory bandwidth, and the two GPUs with higher bandwidth achieve the top performance.
% What is more interesting is that performance trends are different, especially for the HD7970.
% While still being the architecture that achieves the highest performance, it needs bigger input instances to become the best performing architecture, and the margin between this device and the other GPUs is narrower than in the Apertif case.
% The HD7970 is considerably faster than the other accelerators in the Apertif setup because it can exploit its higher memory bandwidth to feed its CUs, and then compute multiple DMs per CU, avoiding expensive memory transfers.
% However, the limited data-reuse of the LOFAR setup implies a decrease in the amount of shared data, forcing the algorithm to compute less DMs per CU and thus increasing the number of transfers from global memory.
% The HD7970 remains faster than the NVIDIA GPUs, but its performance is more affected by the lower AI.
In this setup the GPUs are, on average, 2.5 times faster than the Xeon Phi, but the differences between them are less pronounced than in the Apertif setup.

The line labeled ``\textit{real-time}'', present in both Figures~\ref{fig:PerfApertif} and \ref{fig:PerfLOFAR}, represents the threshold, different for each observational setup, under which the achieved performance would not be enough to dedisperse one second of data in less than one second of computation.
This constraint is fundamental for modern radio telescopes, because their extreme data rate does not allow to store the input for off-line processing.
For all tested input instances, performance achieved by the auto-tuned algorithm is enough to satisfy this constraint, with the only exception represented by the Xeon Phi.
Projecting these results, it can be seen that the GPUs are able to scale to even bigger instances of the problem and still satisfy the real-time constraint.

% \begin{figure}
% \centering
% \includegraphics[width=0.65\columnwidth]{AvgApertif}
% \caption{Average performance and standard deviation, Apertif (higher is better).}
% \label{fig:AvgApertif}
% \end{figure}

% \begin{figure}
% \centering
% \includegraphics[width=0.65\columnwidth]{AvgLOFAR}
% \caption{Average performance and standard deviation, LOFAR (higher is better).}
% \label{fig:AvgLOFAR}
% \end{figure}

\begin{figure}
\centering
\includegraphics[width=0.65\columnwidth]{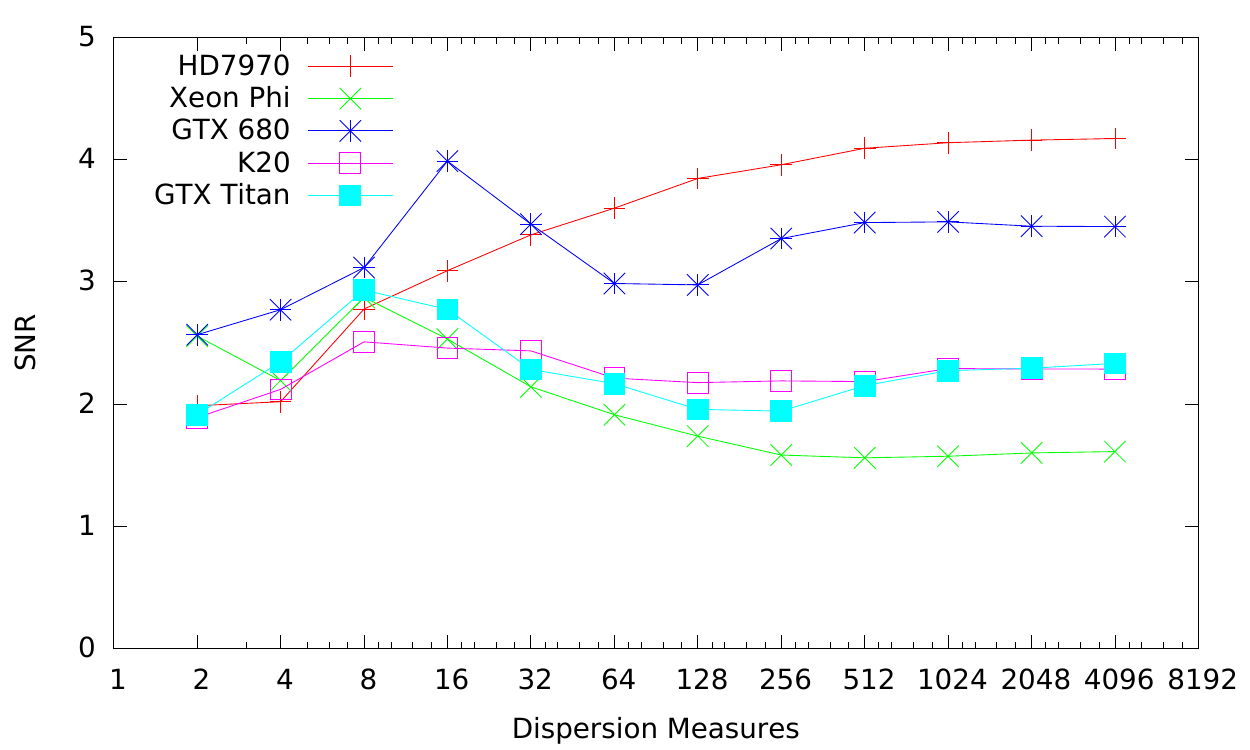}
\caption{Signal-to-noise ratio of the optimum, Apertif.}
\label{fig:SNRApertif}
\end{figure}

\begin{figure}
\centering
\includegraphics[width=0.65\columnwidth]{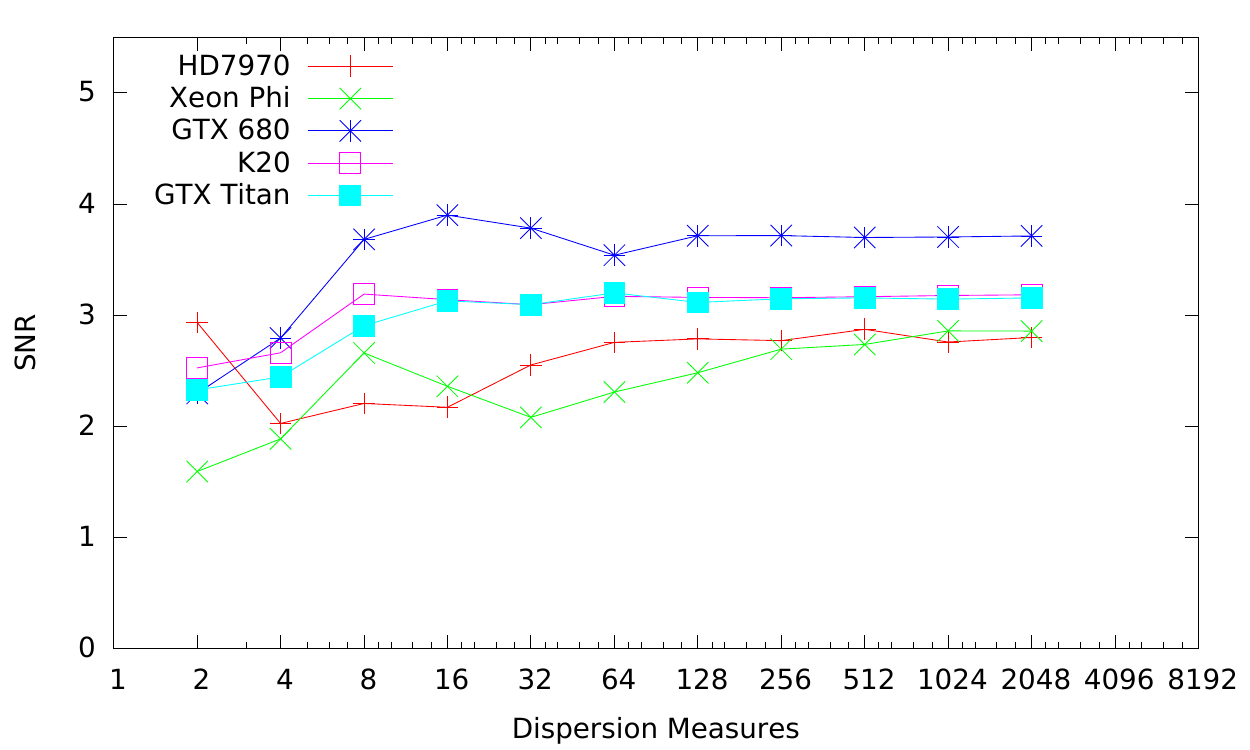}
\caption{Signal-to-noise ratio of the optimum, LOFAR.}
\label{fig:SNRLOFAR}
\end{figure}

\begin{figure}
\centering
\includegraphics[width=0.65\columnwidth]{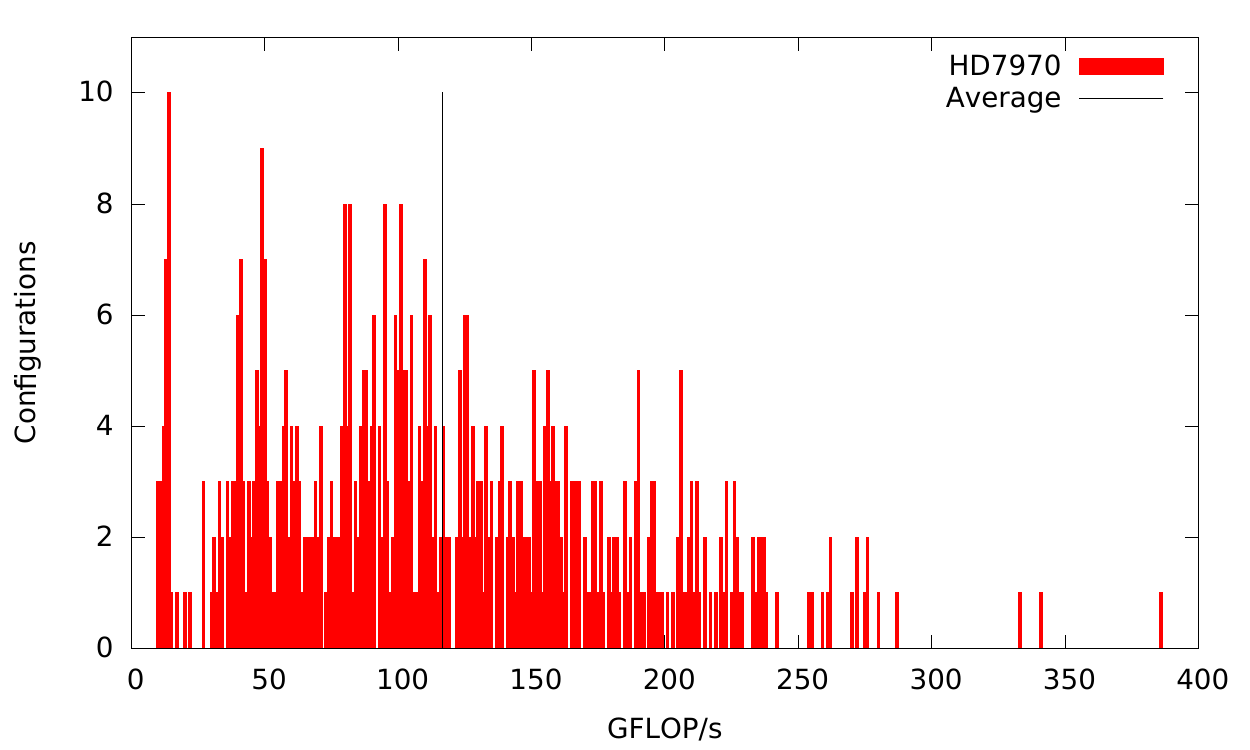}
\caption{Example of performance histogram, Apertif.}
\label{fig:HD7970HistApertif}
\end{figure}

This experiment aimed not just at showing performance and scalability of the tuned dedispersion algorithm, but also at measuring the impact that auto-tuning has on performance.
We believe that it is important to quantify how much faster the tuned algorithm is compared with a generic configuration, and what is the statistical relevance of the optimal performance.
Figures~\ref{fig:SNRApertif} and \ref{fig:SNRLOFAR} present the signal-to-noise ratio of the tuned optimums, i.e. the distance from the average in terms of units of standard deviation.
These results prove that, without using auto-tuning, finding the configuration that provides the best performance would be non-trivial.
Applying Chebyshev's inequality~\cite{abramowitz1964} we can quantify an upper bound on the probability of guessing optimums with these SNRs: in the best case scenario this probability is less than 39\%, while in the worst case it is less than 5\%.
In addition to these results, Figure~\ref{fig:HD7970HistApertif} shows the shape of a typical distribution of the configurations over performance in the optimization space: it can be clearly seen that the optimum lies far from the typical configuration.
In this example, there is exactly one configuration that leads to the best performance; the others perform significantly worse.
We believe that the results of this experiment are an empirical proof of the importance of using auto-tuning, and a good measure of the impact that auto-tuning has on performance.

\subsection{Data-reuse and Performance Limits} % (fold)
\label{sub:data_reuse_and_performance_limits}

\begin{figure}
\centering
\includegraphics[width=0.65\columnwidth]{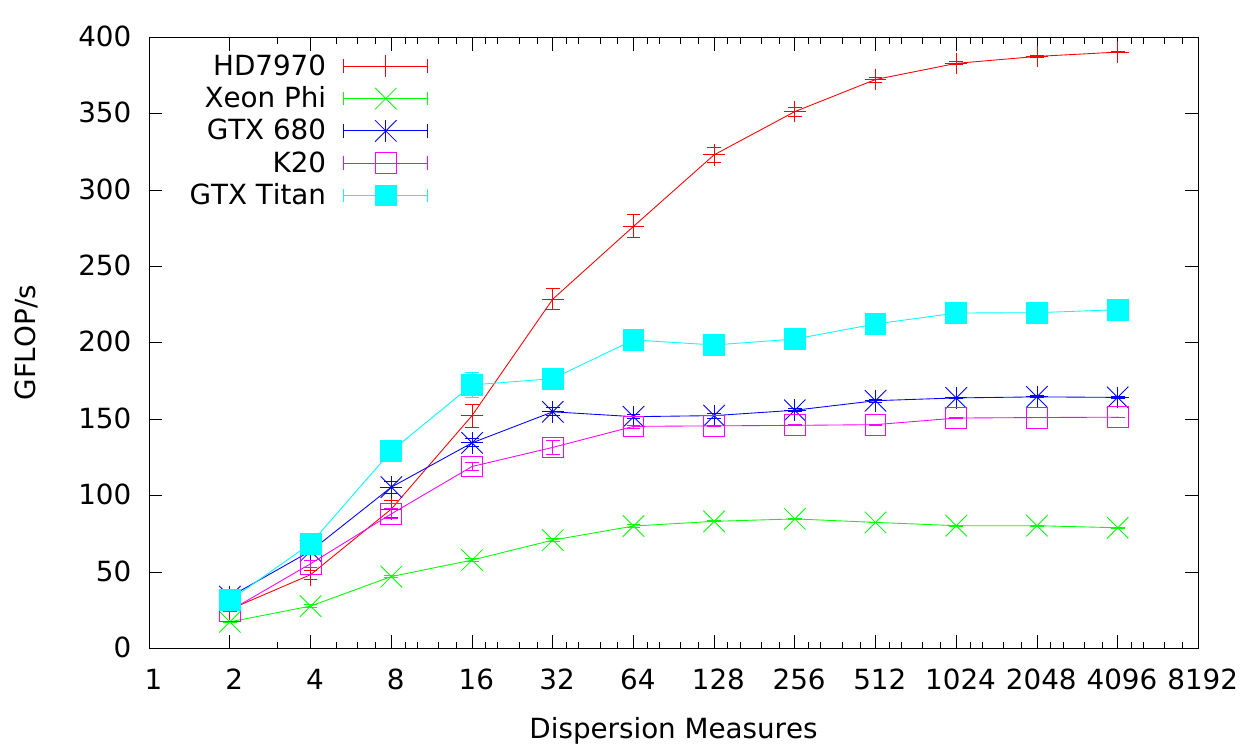}
\caption{Performance in a 0 DM scenario, Apertif (higher is better).}
\label{fig:PerfApertif_0DM}
\end{figure}

\begin{figure}
\centering
\includegraphics[width=0.65\columnwidth]{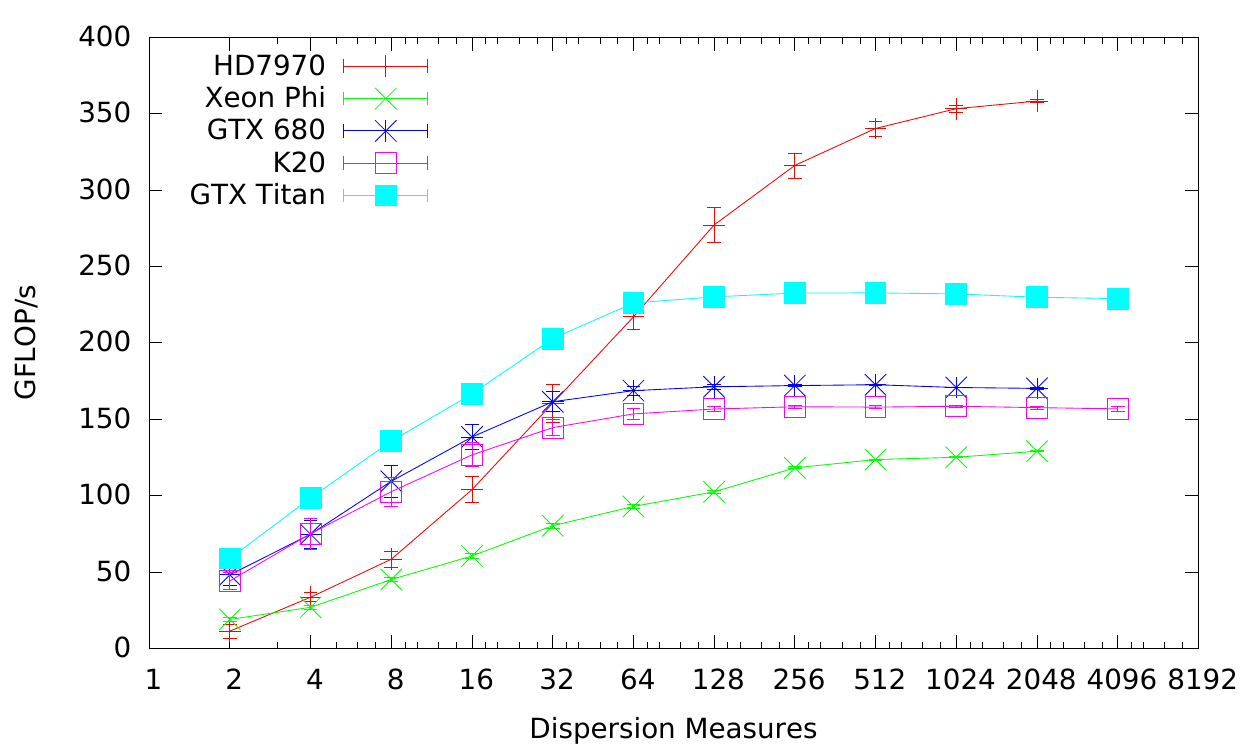}
\caption{Performance in a 0 DM scenario, LOFAR (higher is better).}
\label{fig:PerfLOFAR_0DM}
\end{figure}

%In this section we present and analyze the results of the experiment described in Section~\ref{sub:exp_data_reuse_and_performance_limits}.
In this section we present and analyze the results of the experiment described in Section~\ref{sub:exp_data_reuse_and_performance_limits}, i.e. we tune and measure the performance of dedispersion using the same value, zero, for all DMs.
% In this way we produce an artificial scenario in which it is theoretically possible to achieve perfect data-reuse.
Figures~\ref{fig:PerfApertif_0DM} and \ref{fig:PerfLOFAR_0DM} present the results for the experiment in the Apertif and LOFAR setup, respectively.
In the case of Apertif, the difference between these results and the ones in Figure~\ref{fig:PerfApertif} are negligible, both in terms of scalability and achieved performance.
However, results for the LOFAR setup are clearly different: the performance results are higher and in line with the measurements of the Apertif setup.

The reason of this sudden change is that in this experiment both setups, even if maintaining their differences in terms of computational requirements, expose a theoretically perfect data-reuse to the algorithm, so their difference in performance is reduced.
On the one hand, the fact that performance for the Apertif setup does not change between a real and a perfect setup can be explained by the fact that reuse is already maximized for the real hardware that we tested, and just exposing more data-reuse does not help increasing performance because of hardware limitations.
On the other hand, performance for the LOFAR setup does change, because the increased data-reuse is exploited until the hardware is saturated.

To summarize, with this experiment we once more showed that the observational setup does affect performance.
In particular, what ultimately determines performance is the amount of data-reuse that the observational setup presents to the algorithm, and this is a function of frequencies and DM values.
Data-reuse is so important that, if the observational setup does not expose enough of it, the algorithm is unable to achieve its potential maximum performance.
However, even when perfect data-reuse would make it possible to achieve a theoretically unbounded AI, thus making the algorithm compute-bound, limitations of real hardware do not permit to practically achieve this result.
We therefore conclude that dedispersion is memory-bound for every practical and real scenario.

% subsection data_reuse_and_performance_limits (end)

\subsection{Discussion} % (fold)
\label{sub:discussion}

\begin{figure}
\centering
\includegraphics[width=0.65\columnwidth]{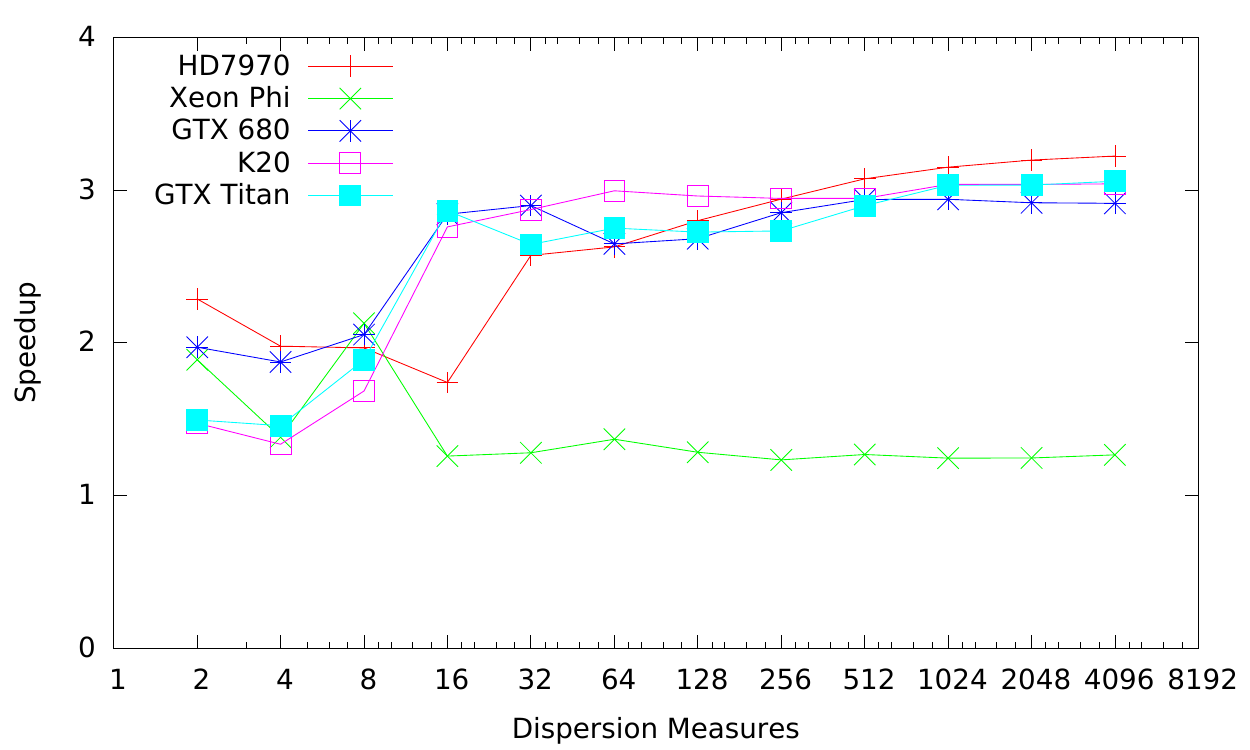}
\caption{Speedup over fixed configuration, Apertif (higher is better).}
\label{fig:SpeedupFixedApertif}
\end{figure}

\begin{figure}
\centering
\includegraphics[width=0.65\columnwidth]{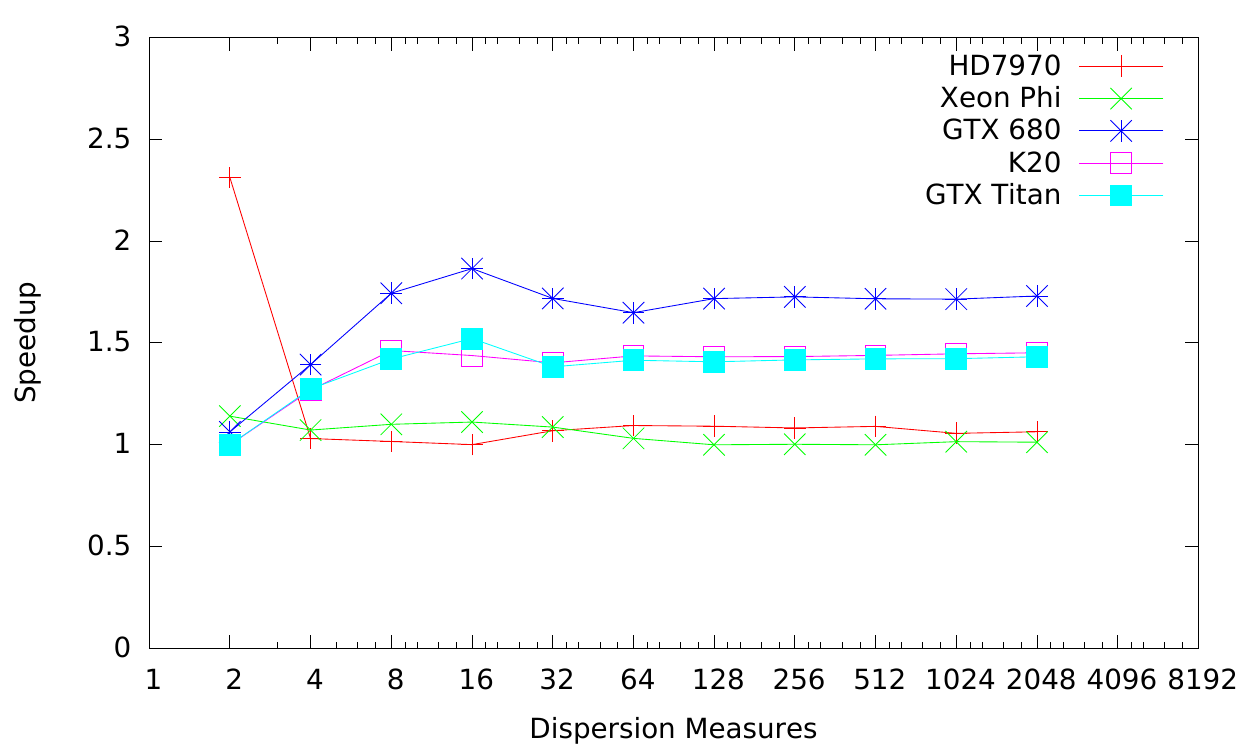}
\caption{Speedup over fixed configuration, LOFAR (higher is better).}
\label{fig:SpeedupFixedLOFAR}
\end{figure}

In this section we present some results that are complementary to the previous experiments, and discuss some more general findings.
We start by comparing the optimal performance of our auto-tuned algorithm to the performance of the \emph{best possible} manually optimized version.
This manually optimized version uses a ``\textit{fixed}'' configuration, i.e. it uses the configuration that, working on all input instances, maximizes the sum of achieved GFLOP/s. We find the \emph{best possible} fixed version with auto-tuning.
This configuration is different for each accelerator and observational setup.
Identifying a single fixed configuration that works on all accelerators and observational setups is possible, but performance would be too low to provide a fair comparison.
% Figures~\ref{fig:FixedApertif} and \ref{fig:FixedLOFAR} show the performance achieved by these fixed configurations for Apertif and LOFAR, respectively.
Figure~\ref{fig:SpeedupFixedApertif} and \ref{fig:SpeedupFixedLOFAR} show the speedup of the auto-tuned algorithm over fixed configurations.

% \begin{figure}
% \centering
% \includegraphics[width=0.65\columnwidth]{FixedApertif}
% \caption{Performance of best non tuned dedispersion, Apertif (higher is better).}
% \label{fig:FixedApertif}
% \end{figure}

% \begin{table}
% \centering
% \begin{tabular}{ | l | r | r | r | }
% \hline
% \textbf{Platform} & \textbf{Min} & \textbf{Average} & \textbf{Max} \\
% \hline
% AMD HD7970 & 1.7 & \textbf{2.6} & \textbf{3.2} \\
% \hline
% Intel Xeon Phi 5110P & 1.0 & 1.4 & 1.6 \\
% \hline
% % NVIDIA GTX 480 & 1.6 & 2.3 & 2.6 \\
% % \hline
% NVIDIA GTX 680 & \textbf{1.9} & \textbf{2.6} & 2.9 \\
% \hline
% NVIDIA K20 & 1.3 & \textbf{2.6} & 3.0 \\
% \hline
% NVIDIA GTX Titan & 1.4 & 2.5 & 3.0 \\
% \hline
% \end{tabular}
% \caption{Speedup of the optimum against non-tuned, Apertif.}
% \label{tab:SpeedupFixedApertif}
% \end{table}

For Apertif, we see that the tuned optimums are 3 times faster than fixed configurations for all GPUs, while the gain in performance for the Xeon Phi is less pronounced.
This difference in performance between tuned and fixed configurations can be seen also in the LOFAR setup, but it is smaller than for Apertif.
In this case, the NVIDIA GPUs still gain a 50\% in performance by auto-tuning, but the HD7970 and Xeon Phi tuned configurations are only slightly faster than the fixed ones.
This is because, as we noted in Section~\ref{sub:auto_tuning}, the smaller optimization space of the LOFAR setup makes the optimum more stable, thus making it easier to manually tune the algorithm.
We believe that these results are further evidence of the importance that auto-tuning has in achieving high performance.

\begin{figure}
\centering
\includegraphics[width=0.65\columnwidth]{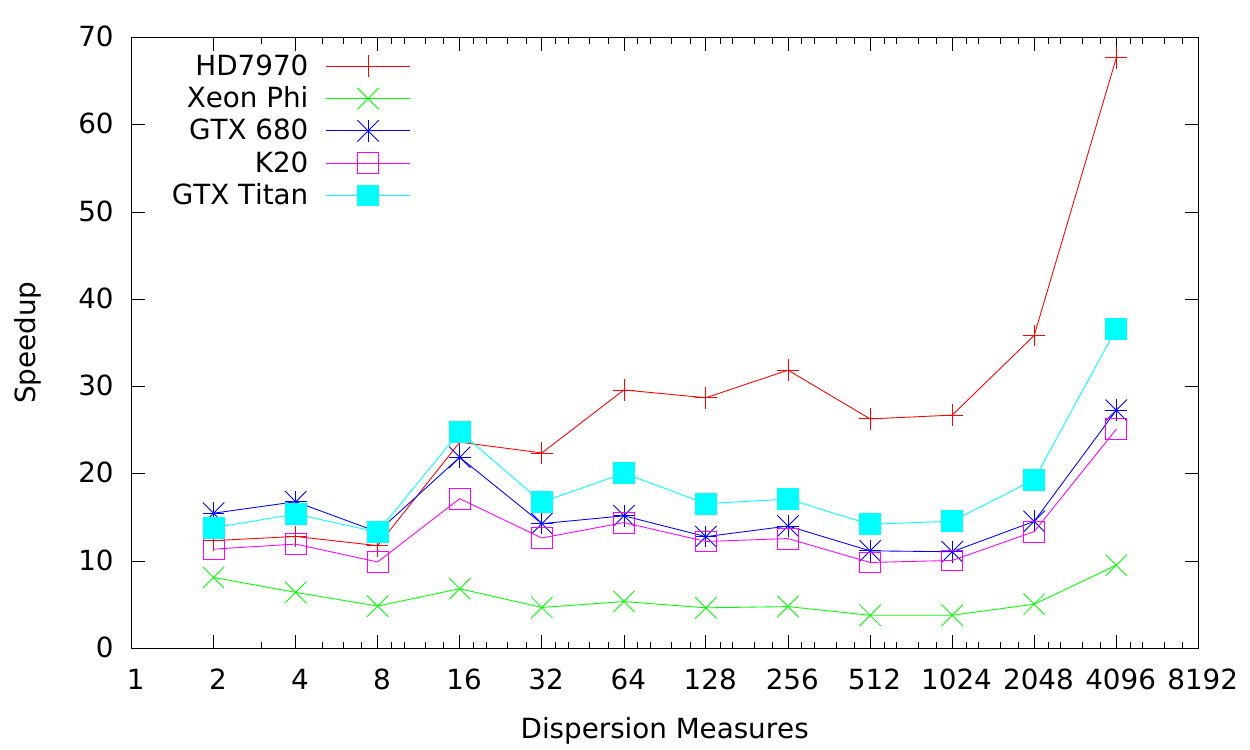}
\caption{Speedup over a CPU implementation, Apertif (higher is better).}
\label{fig:SpeedupApertif}
\end{figure}

\begin{figure}
\centering
\includegraphics[width=0.65\columnwidth]{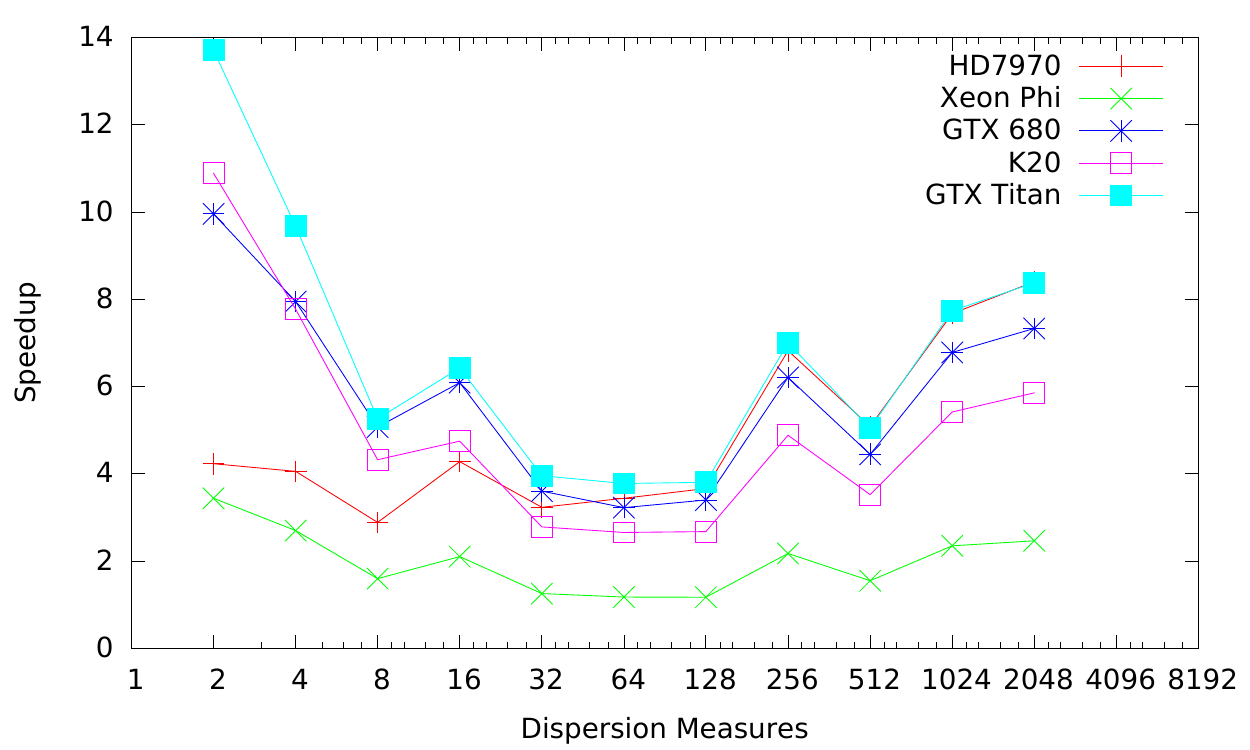}
\caption{Speedup over a CPU implementation, LOFAR (higher is better).}
\label{fig:SpeedupLOFAR}
\end{figure}

We also believe that, currently and in the foreseeable future, many-core accelerators are necessary to achieve high performance for dedispersion.
To provide additional strength to this claim, we compare the performance of our tuned many-core algorithm with an optimized CPU version.
This CPU version of the algorithm is parallelized using OpenMP, with different threads computing different DM values and blocks of time samples.
Chunks of 8 time samples are computed at once using Intel's Advanced Vector Extensions (AVX).
The CPU used to execute the code is the Intel Xeon E5-2620; all the experimental parameters are the same as described in Section~\ref{sec:experimental_setup} except for the used compiler, which is version 13.1.1 of the Intel C++ Compiler (icc).
The speedups over this CPU implementation are presented in Figures~\ref{fig:SpeedupApertif} and \ref{fig:SpeedupLOFAR}.
These results show that our OpenCL dedispersion, running on many-core accelerators, is considerably faster than the CPU implementation in both observational setups.
% Performance comparisons are presented in Figure~\ref{fig:CPUCompApertif} and \ref{fig:CPUCompLOFAR} for Apertif and LOFAR, respectively.

% \begin{figure}
% \centering
% \includegraphics[width=0.65\columnwidth]{CPUComparisonApertif}
% \caption{Performance comparison with a CPU, Apertif (higher is better).}
% \label{fig:CPUCompApertif}
% \end{figure}

% \begin{figure}
% \centering
% \includegraphics[width=0.65\columnwidth]{CPUComparisonLOFAR}
% \caption{Performance comparison with a CPU, LOFAR (higher is better).}
% \label{fig:CPUCompLOFAR}
% \end{figure}

% The only exception is the Xeon Phi, that in the LOFAR setup is, on average, slower than the CPU.
% In this case, the CPU implementation is faster than the many-core one, because with less available data-reuse, the discriminant between the two Intel devices becomes vectorization, and the manual vectorization of the CPU code results more efficient than the one automatically provided by the OpenCL compiler.
% Another trend that we observe is that traditional CPUs, with their limited number of cores, do not scale with the number of DMs to compute, and performance drops after reaching a peak, instead of stabilizing; this result is in accordance with the experiments of Armour et al. in \cite{armour2012}.
% Tables~\ref{tab:SpeedupCPUApertif} and \ref{tab:SpeedupCPULOFAR} contains the achieved speedups over the CPU implementation for the two setups.
% On average, GPUs perform 13--27 and 4--5 times better than the CPU in the Apertif and LOFAR setup, respectively.

With regard to the performance achieved using OpenCL on the Xeon Phi, we believe that in future a better OpenCL implementation for the Phi will certainly increase its performance, and we hope that dedispersion will be able to benefit from the high memory bandwidth of this accelerator.
The aim of future investigations is to tune an OpenMP implementation of the algorithm on the Xeon Phi, and compare its performance with OpenCL.
% However, we do not believe that SIMD architectures are an ideal match for this problem.
% A different implementation, specifically optimized for this architecture, may have performed better, but given that portability is one of the advantages of using OpenCL \cite{du2012}, we do not believe that writing manually optimized versions of the same algorithm for every architecture can be considered a long-term solution.
% Instead, we propose to use auto-tuning to complement OpenCL, providing both code and performance portability.
Furthermore, we also believe the K20 to be a poor match for a memory-bound algorithm like dedispersion, because it does not have enough memory bandwidth to feed its compute elements and keep them busy.

% \begin{table}
% \centering
% \begin{tabular}{ | l | r | r | r | }
% \hline
% \textbf{Platform} & \textbf{Min} & \textbf{Average} & \textbf{Max} \\
% \hline
% AMD HD7970 & \textbf{11.8} & \textbf{27.4} & \textbf{67.7} \\
% \hline
% Intel Xeon Phi 5110P & 3.6 & 5.2 & 9.3 \\
% \hline
% NVIDIA GTX 480 & 10.4 & 14.1 & 26.7 \\
% \hline
% NVIDIA GTX 680 & 11.1 & 15.6 & 27.3 \\
% \hline
% NVIDIA K20 & 9.8 & 13.3 & 25.1 \\
% \hline
% \end{tabular}
% \caption{Speedup over an optimized CPU implementation, Apertif.}
% \label{tab:SpeedupCPUApertif}
% \end{table}

% \begin{table}
% \centering
% \begin{tabular}{ | l | r | r | r | }
% \hline
% \textbf{Platform} & \textbf{Min} & \textbf{Average} & \textbf{Max} \\
% \hline
% AMD HD7970 & 2.9 & 4.8 & 8.4 \\
% \hline
% Intel Xeon Phi 5110P & 1.2 & 1.8 & 2.6 \\
% \hline
% NVIDIA GTX 480 & 2.9 & 5.2 & \textbf{11.2} \\
% \hline
% NVIDIA GTX 680 & \textbf{3.2} & \textbf{5.7} & 9.9 \\
% \hline
% NVIDIA K20 & 2.6 & 4.9 & 10.9 \\
% \hline
% \end{tabular}
% \caption{Speedup over an optimized CPU implementation, LOFAR.}
% \label{tab:SpeedupCPULOFAR}
% \end{table}

 With the performance results of this work, we can also compute the number of accelerators that would be necessary to implement real-time dedispersion for Apertif.
Apertif will need to dedisperse in real-time 2,000 DMs, and do this for 450 different beams.
Using our best performing accelerator, the AMD HD7970, it is possible to dedisperse 2,000 DMs in 0.106 seconds; combining 9 beams per GPU it would still be theoretically possible to dedisperse one second of data in real-time, with enough available memory to store both the input and the dedispersed time-series.
Therefore, dedispersion for Apertif could be implemented today with just 50 GPUs, instead of the 1,800 CPUs that would be necessary otherwise.

% subsection discussion (end)

% section results_and_discussion (end)

\section{Related Work} % (fold)
\label{sec:related_works}

In the literature, auto-tuning is considered a viable technique to achieve performance that is both high and portable.
In particular, the authors of \cite{li2009} show that it is possible to use auto-tuning to improve performance of even highly-tuned algorithms.
Even more relevant to our work is \cite{du2012}, and we agree with the authors of \cite{du2012} that auto-tuning can be used as a performance portability tool, especially with OpenCL.
However, auto-tuning has been used mostly to achieve performance portability between different many-core accelerators, while we also use it to adapt an algorithm to different use-case scenarios.

While there are no previous attempts at auto-tuning dedispersion for many-cores, there are a few previous GPU implementations documented in literature.
The first reference in the literature is \cite{barsdell2010}. In this paper dedispersion is listed as a possible candidate for acceleration, together with other astronomy algorithms.
We agree with the authors of \cite{barsdell2010} that dedispersion is a potentially good candidate for many-core acceleration, because of its inherently parallel structure, but we believe their performance analysis to be too optimistic, and their AI's estimate to be unrealistically high.
In fact, we showed in this paper that dedispersion's AI is low in all realistic scenarios, and that the algorithm is inherently memory-bound.
The same authors implemented, in a follow-up paper~\cite{barsdell2012b}, dedispersion for NVIDIA GPUs, using CUDA as their implementation framework.
However, we do not completely agree with their performance results for two reasons: first, they do not completely exploit data-reuse, and we showed how important data-reuse is for performance, and second, part of their results are not experimental, but derived from performance models.

Another GPU implementation of the dedispersion algorithm is presented in \cite{magro2011}.
Also in this case there is no mentioning of exploiting data-reuse.
In fact, some of the authors of \cite{magro2011} published, shortly after the first work, another short paper \cite{armour2012} in which they affirm that the previous algorithm does not perform well enough because it does not exploit data-reuse.
Unfortunately, this paper does not provide sufficient detail on either the algorithm or on experimental details such as frequencies and time resolution, for us to repeat their experiment.
Therefore, we cannot verify the claimed 50\% of theoretical peak performance.
However, we believe this claim to be unrealistic because dedispersion has an inherently low AI, and it cannot take advantage of fused multiply-adds, which by itself already limits the theoretical upper bound to 50\%.

% section related_works (end)

\section{Conclusions} % (fold)
\label{sec:conclusions}

In this paper, we analyzed dedispersion, a basic radio astronomy algorithm that is used to reconstruct smeared signals, especially when searching for new celestial objects.
We introduced the algorithm and our many-core implementation, and analytically proved that dedispersion is a memory-bound algorithm and that, in any real case scenario, its performance is limited by low AI.
With the experiments presented in this paper, we also demonstrated that by using auto-tuning it is possible to obtain high performance for dedispersion.
Even more important, we showed that auto-tuning makes the algorithm portable between different platforms and different observational setups.
Furthermore, we highlighted how auto-tuning permits to automatically exploit the architectural specificities of different platforms.

Measuring the performance of the tuned algorithm, we verified that it scales linearly with the number of DMs for every tested platform and observational setup.
So far, the most suitable platform to run dedispersion among the ones we tested, is a GPU from AMD, the HD7970.
This GPU performs better than the other accelerators when extensive data-reuse is available, and achieves good performance also in less optimal scenarios, thanks to its high memory bandwidth.
If there is less data-reuse, the GPUs that we tested achieve similar performance, but are still 2--7 times faster than the Intel Xeon Phi.
Although this is partially due to the Xeon Phi's immature OpenCL implementation, we have to conclude that, at the moment, GPUs are better candidates for dedispersion.

Another important contribution was the quantitative evidence of the impact that auto-tuning has on performance.
With our experiments, we showed that the optimal configuration is difficult to find manually and lies far from the average, having an average signal-to-noise ratio of 2--4.
Moreover, we showed that the auto-tuned algorithm is faster than manually tuned versions of the same algorithm on all platforms, and is an order of magnitude faster than an optimized CPU implementation.

Finally, our last contribution was to provide further empirical proof that dedispersion is a memory-bound algorithm, limited by low AI.
In particular, we showed that achievable performance is limited by the amount of data-reuse that dedispersion can exploit, and the available data-reuse is affected by parameters like the DM space and the frequency interval.
We also showed that, even in a perfect scenario where data-reuse is unrealistically high, the performance of dedispersion is limited by the constraints imposed by real hardware, and approaching the theoretical AI bound of the algorithm becomes impossible.

% section conclusions (end)

\bibliographystyle{IEEEtran}
\bibliography{Bibliography}

\end{document}